\documentclass[aps,prl,twocolumn,superscriptaddress,showpacs]{revtex4}

\usepackage{graphicx}

\begin{document}
\title{Measurement of long-range wavefunction correlations in an open microwave billiard}
\author{Y.-H.~Kim}
\author{U.~Kuhl}
\author{H.-J.~St\"ockmann}
\affiliation{Fachbereich Physik der Philipps-Universit\"at
Marburg, Renthof 5, D-35032 Marburg, Germany}
\author{P.W.~Brouwer}
\affiliation{Laboratory of Atomic and Solid State Physics, Cornell
University, Ithaca, New York 14853-2501, USA}
\date{\today}

\begin{abstract}
We investigate the statistical properties of wavefunctions in an open
chaotic cavity. When the number of channels in the openings of the
billiard is increased by varying the frequency, wavefunctions cross
over from real to complex. The distribution of the phase rigidity,
which characterizes the degree to which a wavefunction is complex, and
long-range correlations of intensity and current density are studied
as a function of the number of channels in the openings. All measured
quantities are in perfect agreement with theoretical predictions.
\end{abstract}

\pacs{73.23.-b, 73.23.Ad,05.45.Mt, 72.20.-i} \maketitle


{}From a statistical point of view, eigenvalues and eigenfunctions
of the wave equation in a chaotic billiard are well described by
random matrix theory~\cite{kn:guhr1998}: Depending on the presence
or absence of time-reversal symmetry, they show the same
distribution as eigenvalues and eigenvectors of a large hermitian
matrix with random Gaussian distributed real or complex elements.
Whereas random matrix theory predicts a very characteristic
eigenvalue distribution --- with eigenvalue repulsion and spectral
rigidity~\cite{kn:mehta1991} ---, its predictions for eigenvectors
are rather ``uninteresting'': eigenvector elements are independent
Gaussian distributed real or complex random numbers.

The situation is different for the eigenvectors of a random matrix
that interpolates between the standard ensembles with real and
complex matrix elements. An example of such an interpolating
ensemble is the ``Pandey-Mehta'' Hamiltonian~\cite{kn:pandey1983}
\begin{equation}
  H(\alpha) = H_0 + \alpha H_1, \label{eq:Pandey}
\end{equation}
where $H_0$ and $H_1$ are real and complex random  hermitian
matrices, respectively, and $\alpha$ is a crossover parameter. For
such an interpolating random matrix ensemble the eigenvector
elements have a non-Gaussian
distribution~\cite{kn:sommers1994,kn:falko1994} and acquire
correlations, both between elements of the same
eigenvector~\cite{kn:falko1996,kn:vanlangen1997} and between
different eigenvectors~\cite{kn:adam2002}.

Experimental verification of this eigenvector distribution has
proven problematic because the long-range correlations and the
deviations from Gaussian distributions are only of the order of a
few
percent~\cite{kn:sommers1994,kn:falko1994,kn:falko1996,kn:vanlangen1997,kn:adam2002}.
An additional complication arises from the fact that the crossover
parameter $\alpha$ needs to be fitted to the experiment. These
complications may explain why measurements of wavefunction
distributions in two-dimensional microwave cavities in which
time-reversal symmetry was broken using magneto-optical effects
were inconclusive with respect to the functional form of the
probability distribution and did not reveal long-range
wavefunction correlations~\cite{kn:chung2000}.

An alternative method to observe the real-to-complex crossover is
to consider traveling waves in a billiard that is opened to the
outside world~\cite{kn:pnini1996}. For microwaves, such an open
billiard is obtained by connecting a two-dimensional microwave
cavity to waveguides. The parameter governing the crossover from
real to complex wavefunctions is the total number of channels $N$
in the two waveguides. As was shown by one of the
authors~\cite{bro03}, wavefunctions in this crossover have a
non-Gaussian distribution and long-range correlations, just like
the eigenvectors of the Pandey-Mehta
Hamiltonian~(\ref{eq:Pandey}). The main difference, however, is
that the crossover parameter $N$ is discrete and can be measured
independently. This allows a fit-parameter free comparison of
theory and experiment.

We here report on the first measurement of
such long-range wavefunction correlations in the real-to-complex
crossover.
The basic principles of the experiment can be found in
Ref.~\onlinecite{Kuh00b}. We used a rounded rectangular
cavity\,(21\,cm$\times$\,18\,cm) coupled to two waveguides of
width 3\,cm with a cut-off frequency at $\nu_T=5$\,GHz. To break
the symmetry of the shape of the resonator and to block direct
transport, two half disks with a radius of 3\,cm were placed in
the resonator. To avoid unwanted reflection, absorbers were placed
at the end of the leads. We scanned the billiard on a square grid
of 5\,mm with a movable antenna $A_1$ and measured transmission
$S_{12}$ in the range of 4\,-\,18\,GHz from a fixed antenna $A_2$
in the end of the right lead. The fixed antenna had a metallic
core of diameter 1\,mm and a teflon coating while the probe
antenna $A_1$ was a thin wire of diameter 0.2\,mm to minimize the
leakage current. The lengths of the antenna $A_1$ and antenna
$A_2$ were 4 and 5\,mm respectively.

\begin{figure}
\parbox{0.99\hsize}{
\hspace*{-.7cm}
\includegraphics*[width=0.55\hsize]{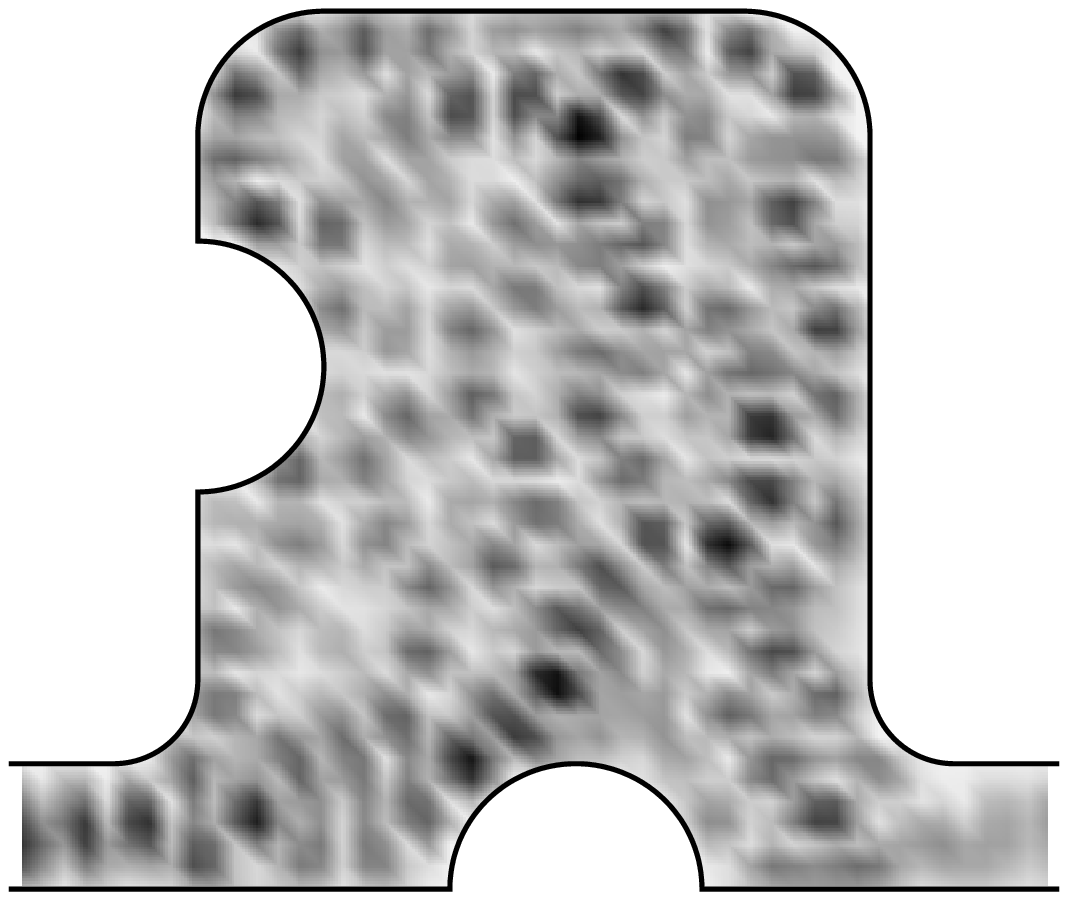}
\hspace*{-.8cm}
\includegraphics*[width=0.55\hsize]{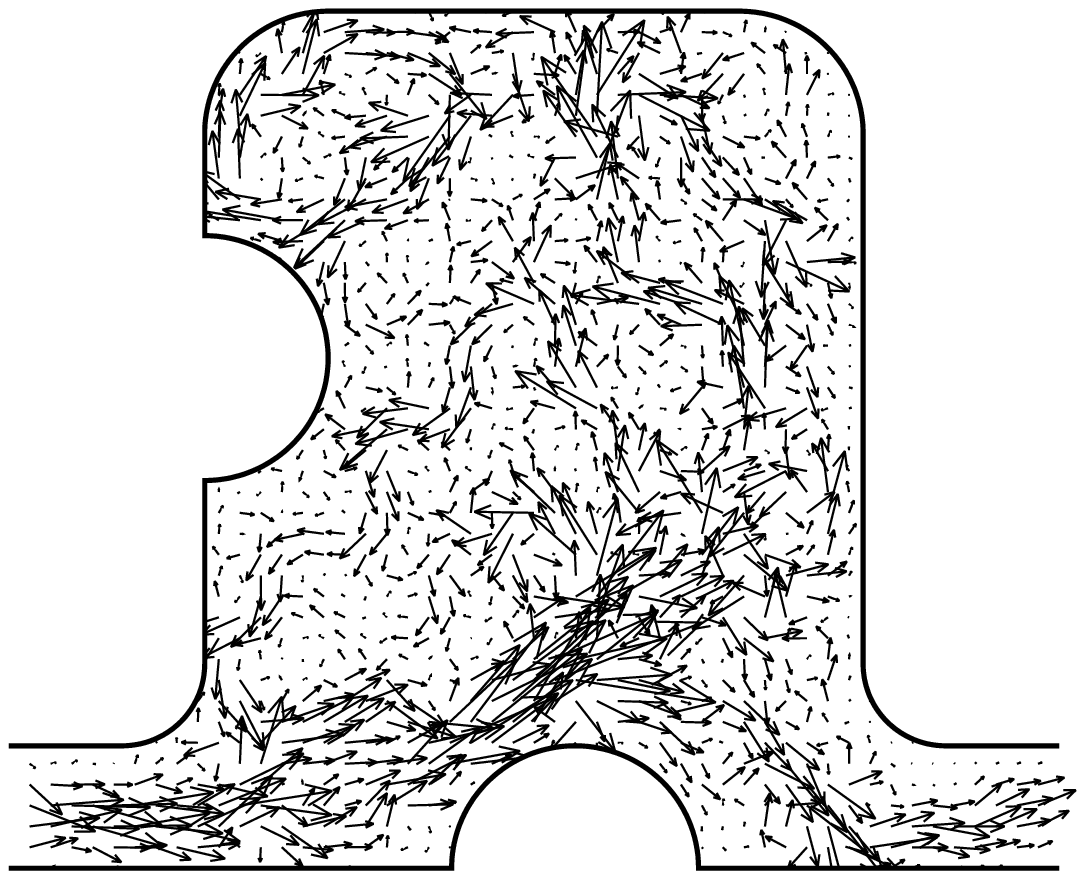}}
\caption{\label{fg:Ij} Left panel: Grey scale plot of measured
intensity~$|\psi|^2$ at $\nu=12.015$\, GHz. Black corresponds to
maximum intensity. Right panel: Measured current density for the
same wave function . Arrow lengths indicate the magnitude of the
Poynting vector. }
\end{figure}

For microwave frequencies $\nu <\nu_{\rm{max}}= c/2d=18.75\,$GHz,
where $c$ is the velocity of light and $d$ is  the resonator
height, the billiard is quasi-two-dimensional. In this regime
there is an exact correspondence between electrodynamics and
quantum mechanics, where the component of the electric field
perpendicular to the plane of the microwave billiard $E_z$
corresponds to the quantum-mechanical wave function $\psi$. We
normalize the wavefunction $\psi$ such that
$  \int d\mathbf{r} |\psi(\mathbf{r})|^2 = 1$.
Then the square of the electric field and the Poynting vector
map to the normalized ``intensity'' and ``current density'',
respectively,
\begin{equation}
  I(\mathbf{r}) = A |\psi(\mathbf{r})|^2,\ \
  \mathbf{j}(\mathbf{r}) = \frac{A}{k}{\rm
  Im}[\psi^*(\mathbf{r})\nabla\psi(\mathbf{r})],
  \label{current}
\end{equation}
where $A$ is the area of the billiard and $k$ the wavenumber.
Figure\,\ref{fg:Ij} shows typical intensity and current patterns
thus obtained.

The correspondence between the perpendicular component of the
electric field $E_z$ and the wavefunction $\psi$ has been used
previously to study the spatial
distributions and correlation functions of currents and vortices
in open billiards~\cite{Bar02, Kim03b}. In the present work we
study spatial correlation functions of the \emph{squares} of
intensities and currents. The statistical average is taken by
averaging over both position and frequency. It is for these
quantities and for this full ensemble average that long-range
correlations were predicted~\cite{bro03}. Before we describe the
experimental results, we briefly summarize the conclusions of
Ref.~\onlinecite{bro03}.

In Ref.~\onlinecite{bro03} the wavefunction distribution was
described as the convolution of a Gaussian distribution with
correlated real and imaginary parts and that of a single complex
number $\rho$, the dot product of the wavefunction $\psi$ and its
time reversed,
\begin{equation}\label{rho}
 \rho= \int d\mathbf{r}\,\psi(\mathbf{r})^2.
\end{equation}
The absolute value $|\rho|^2$ is known as
the ``phase rigidity'' of $\psi$
\cite{kn:vanlangen1997}. The Gaussian wavefunction distribution at a
fixed value of $\rho$ implies a generalized Porter-Thomas distribution
for the intensity,
\begin{equation}
  P_{\rho}(I)= \frac{1}{\sqrt{1 - |\rho|^2}}
  \exp \left[- \frac{I}{1 - |\rho|^2} \right]
  I_0 \left[ \frac{|\rho|I}{1 - |\rho|^2} \right],
  \label{eq:PPT}
\end{equation}
so that the full intensity distribution is obtained by convolution
of Eq.~(\ref{eq:PPT}) and the distribution $p(\rho)$ of $\rho$,
\begin{equation}
  P(I) = \int d\rho\,p(\rho)\,P_{\rho}(I).
\end{equation}
The distribution $p(\rho)$ was calculated in
Ref.~\onlinecite{bro03} using random matrix theory.

In order to explain the origin of long-range correlations of
intensity and current density in an open chaotic billiard, we
consider the joint distribution of intensities at points
$\mathbf{r}$ and $\mathbf{r}'$ with separation $k|\mathbf{r} -
\mathbf{r'}| \gg 1$,
\begin{equation}
  P[I(\mathbf{r}),I(\mathbf{r}')] =
  \int d\rho\,p(\rho)\,P_{\rho}[I(\mathbf{r})]\,P_{\rho}[I(\mathbf{r'})].
  \label{eq:Pdist}
\end{equation}
For an open billiard, $\rho$ has a nontrivial distribution, hence
the long-range correlations of $P[I(\mathbf{r}),I(\mathbf{r}')]$.
Whereas random matrix theory predicts the long-range correlations
of intensities through Eq.~(\ref{eq:Pdist}), it cannot alone
predict the long-range correlations of current densities and the
precise dependence of these correlators on the separation
$|\mathbf{r} - \mathbf{r'}|$ for $k|\mathbf{r} - \mathbf{r'}|$ of
order unity. However, as shown in Ref.~\onlinecite{bro03}, the
latter can be obtained from the random matrix result by making use
of Berry's ansatz~\cite{kn:berry1977}, which expresses $\psi$ as a
random sum over plane waves,
\begin{equation}
  \psi(\mathbf{r}) = \sum_{\mathbf{k}} a(\mathbf{k})
  e^{i \mathbf{k} \cdot \mathbf{r}}.
  \label{eq:sum}
\end{equation}
Here the plane wave amplitudes $a(\mathbf{k})$ have a Gaussian
distribution with zero mean and with variance
\begin{equation}
  \langle a(\mathbf{k}) a(-\mathbf{k}) \rangle =  \rho \langle a(\mathbf{k}) a^*(\mathbf{k}) \rangle,
  \label{eq:avar}
\end{equation}
where $\rho$ is the (random) phase rigidity of $\psi$. Performing
the ensemble average using Eqs.\,(\ref{eq:sum}) and
(\ref{eq:avar}), correlators of intensity and current density are
then expressed in terms of moments of the phase rigidity
$|\rho|^2$. Long range correlations are found for correlators
involving the square of the intensity and the current density,
\begin{eqnarray}
\langle I(\mathbf{r})^2I(\mathbf{r}')^2 \rangle_c &=& \mbox{var}\,
    |\rho|^2+4   f^2
\langle4+13|\rho|^2+|\rho|^4\rangle \nonumber\\
   && \mbox{} + 4 f^4
  \langle 1+4|\rho|^2+|\rho|^4\rangle, \label{C_I2I2} \nonumber \\
  \langle I(\mathbf{r})^2 J(\mathbf{r}')^2 \rangle_c&=&-\frac{1}{4}
  \mbox{var}\, |\rho|^2 +
  f^4
  \langle 2-|\rho|^2-|\rho|^4\rangle,  \label{C_I2J2}  \nonumber \\
  \langle J(\mathbf{r})^2 J(\mathbf{r}')^2 \rangle_c&=&\frac{1}{4}
  \mbox{var}\,
  |\rho|^2+\frac{1}{2} f^2
  \langle1-2|\rho|^2+|\rho|^4\rangle \nonumber\\
 && \mbox{} +\frac{1}{4} f^4
  \langle 3-5|\rho|^2+2|\rho|^4\rangle.\label{C_J2J2}
\end{eqnarray}
Here $f = J_0(k|\mathbf{r}-\mathbf{r'}|)$, $J_0$ being the Bessel
function, and the subscript ``$c$'' refers to the connected
correlator, $\langle A B \rangle_c = \langle A B \rangle - \langle
A \rangle \langle B \rangle$. The relevant moments of the phase
rigidity $|\rho|^2$ were calculated in Ref.~\onlinecite{bro03}:
For $N = 2$, $4$, and $6$, one has $\langle |\rho|^2 \rangle =
0.7268$, $0.5014$, and $0.3918$, and $\langle |\rho|^4 \rangle =
0.6064$, $0.3285$, and $0.2155$, respectively.

\begin{figure}
\parbox[h]{0.99\hsize}{
\includegraphics*[width=0.8\hsize]{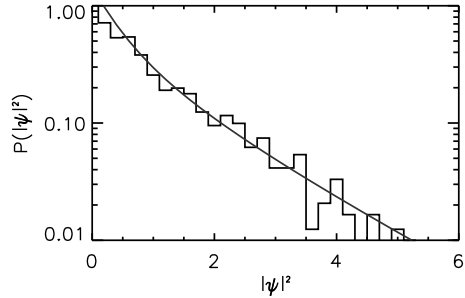}}
\caption{\label{fg:PI} Intensity distribution for the  wave
function at $\nu=12.015$\,GHz. Theoretical values from
Eq.~(\ref{eq:PPT}) are plotted with solid line.}
\end{figure}
We now describe the measured wavefunction distributions. We first
discuss wavefunction distributions measured at a fixed frequency,
and compare to the theory for the corresponding fixed value of
$\rho$. A statistical distribution at a fixed frequency is
obtained by varying the position of the antenna only. The phase
rigidity $|\rho|^2$ can be measured independently using
Eq.~(\ref{rho}). Fig.~\ref{fg:PI} shows the intensity distribution
for the intensity pattern shown in the left panel of
Fig.~\ref{fg:Ij} together with the theory of Eq.~(\ref{eq:PPT}),
using the measured value of $|\rho|^2=0.5202$.
As was discussed in Ref.~\onlinecite{Bar02}, there are frequency
regimes where the leakage to the probe antenna becomes intolerably
high, either due to the fact that the transport through the cavity
is small, or due to some strongly scarred wave functions. In all
such cases there were strong deviations from the generalized
Porter-Thomas behavior described by Eq.~(\ref{eq:PPT}). We
therefore only used frequency regimes where $P(|\psi|^2)$ was in
agreement with theory on a confidence level of 90 percent. To have
a  well defined number of transversal modes within each waveguide,
we investigate the frequency regimes 5\,-\,9.5, 10\,-\,14.5,
15\,-\,18\,GHz, corresponding to a total number of channels
$N$=2,\,4,\,6, respectively ({\em i.e.}, $1$, $2$, and $3$
propagating modes in each waveguide)
\begin{figure}
\begin{center}
\includegraphics[width=0.8\hsize]{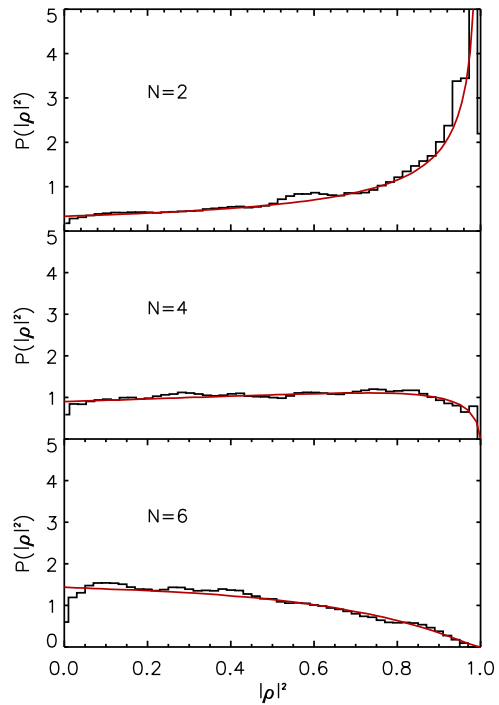}
\caption{\label{fg:rho}Phase rigidity distribution $P(|\rho|^2)$
for different values of the total number of channels $N$. Solid
curves indicate the theory of Ref.~\protect\onlinecite{bro03}.}
\end{center}
\end{figure}
%
\begin{figure}
\begin{center}
\includegraphics[width=0.85\hsize]{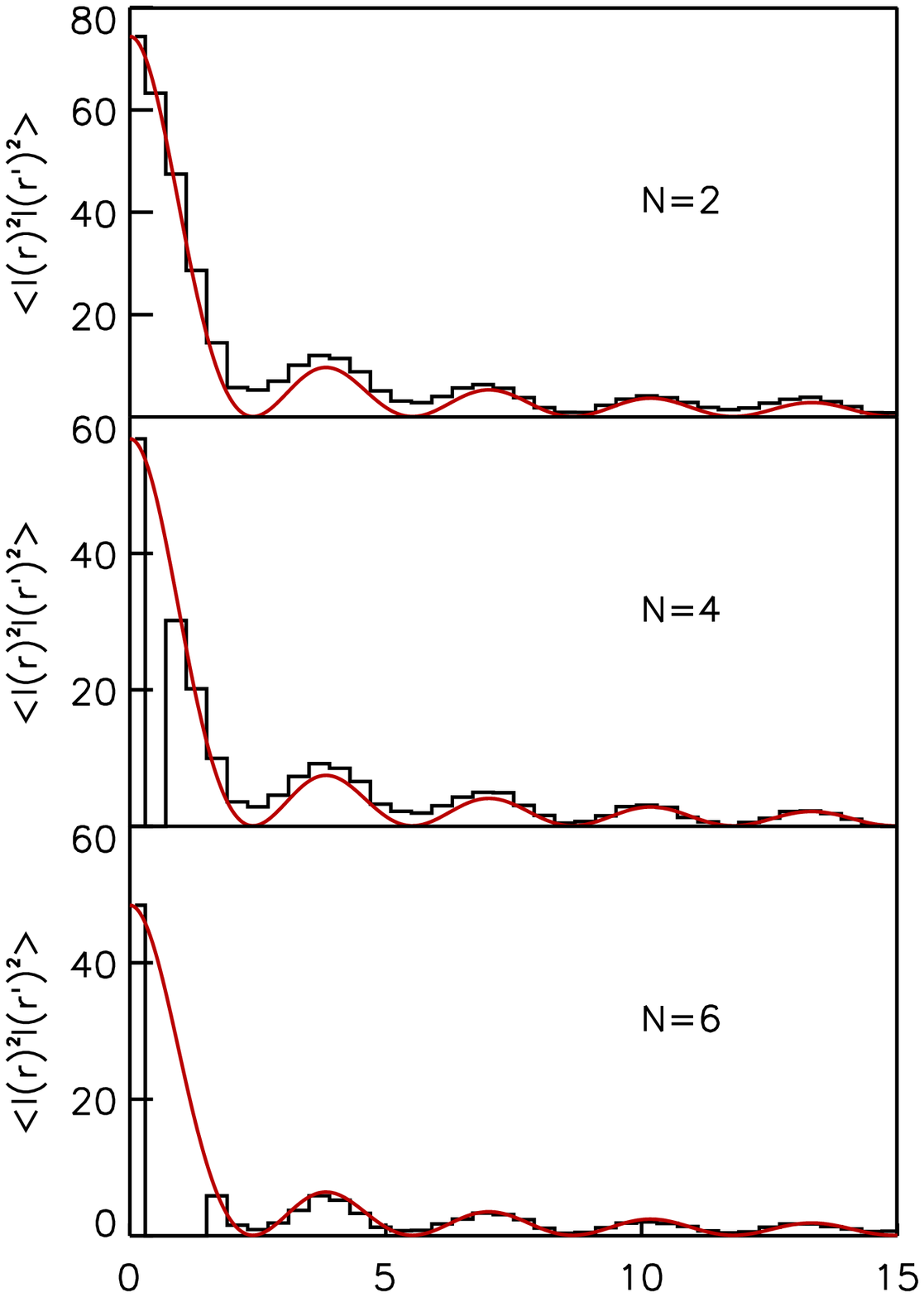}
\caption{\label{fg:cii}Connected correlator of squared intensity
$\langle I(\mathbf{r})^2I(\mathbf{r'})^2\rangle_c$ versus
$k|\mathbf{r}-\mathbf{r'}|$ for different values of the total
number of channels $N$.}
\end{center}
\end{figure}
\begin{figure}
\begin{center}
\includegraphics[width=0.85\hsize]{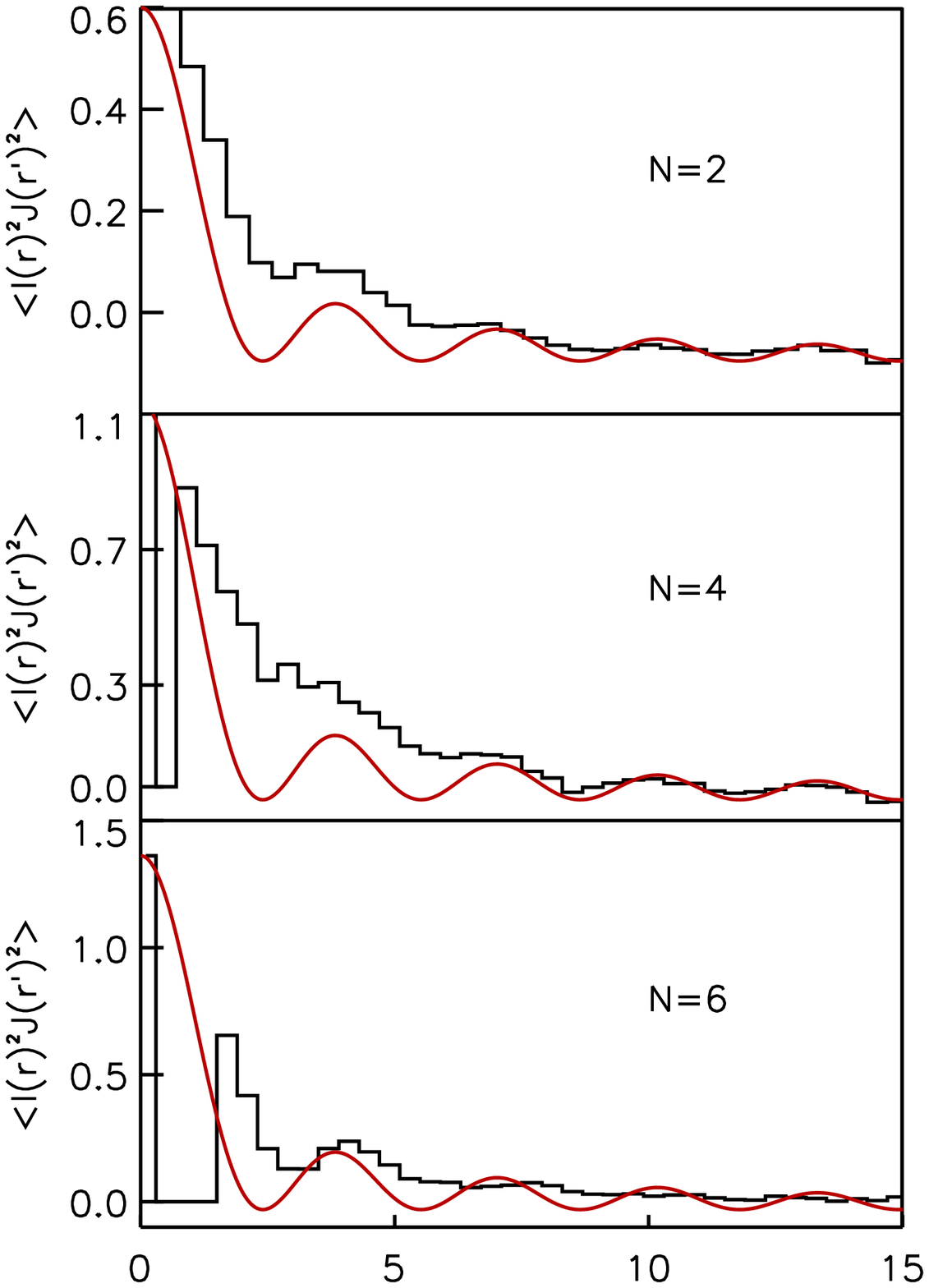}
\caption{\label{fg:cij} Connected correlator  of squared intensity
and squared current density $\langle
I(\mathbf{r})^2J(\mathbf{r'})^2\rangle_c$ versus
$k|\mathbf{r}-\mathbf{r'}|$ , for different values of the total
number of channels $N$.}
\end{center}
\end{figure}
\begin{figure}
\begin{center}
\includegraphics[width=0.85\hsize]{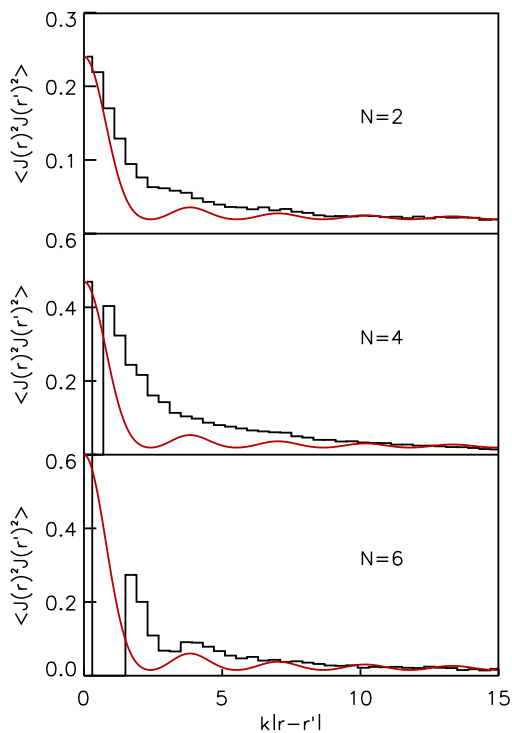}
\caption{\label{fg:cjj} Connected correlator of squared current
density $\langle J(\mathbf{r})^2 J(\mathbf{r'})^2\rangle_c$ versus
 $k|\mathbf{r}-\mathbf{r'}|$  for different values of the
total number of channels $N$.}
\end{center}
\end{figure}

We now describe our results for a full ensemble average, in which
both the position of the detector antenna and the frequency are
varied. Fig.~\ref{fg:rho} shows the measured phase rigidity
distribution $P(|\rho|^2)$ together with the theory of
Ref.~\onlinecite{bro03}, for $N=2$, $4$, and $6$. Good agreement
is found between experiment and theory, especially as there is no
free parameter. Figures~\ref{fg:cii},~\ref{fg:cij}, and
\ref{fg:cjj} show measurement and theoretical prediction for the
correlation functions of the squared intensity and the squared
current density at positions $\mathbf{r}$ and $\mathbf{r}'$. Since
these correlation functions depend on the positions $\mathbf{r}$
and $\mathbf{r}'$ through the combination
$k|\mathbf{r}-\mathbf{r'}|$ only, results from different frequency
regimes can be superimposed by a proper scaling. In our analysis,
we selected frequency windows for averaging of
$\Delta\nu=0.3$\,GHz guaranteeing that $\Delta\nu\ll {c}/{2\nu
L}$, where $L$ is the billiard size and $c$ the velocity of wave
propagation. The gaps in the  $N=4,~6$ histograms for small values
of $k|\mathbf{r}-\mathbf{r'}|$ reflect the spatial resolution
limited to 5\,mm due to the chosen grid size. For the long-range
correlations, we observe excellent agreement between experiment
and theory. The short-range oscillations predicted by theory are
suppressed to a large extent in the experiment for the correlators
that involve the current density, however. This, again, is a
consequence of the limited resolution. We would like to stress,
however, that the asymptotic values of all correlations in the
limit $k|\mathbf{r}-\mathbf{r'}|\rightarrow \infty$ is different
from zero and in perfect agreement with the predictions from
Eq.~(\ref{C_J2J2}).

Long-range correlations have also been observed in the
transmission of microwaves~\cite{Seb02} and visible
light~\cite{Emi03} through two- and three-dimensional disordered
media, in which wave propagation is diffusive. There are important
differences between the long-range correlations observed in Refs.\
\onlinecite{Seb02,Emi03} and those in
 ballistic systems, which are reported here.
First, in Refs.~\onlinecite{Seb02,Emi03}, long-range correlations
appear already for the intensity autocorrelation function, whereas
we find long-range correlations for correlators of second and
higher moments only. Second, in Refs.~\onlinecite{Seb02,Emi03},
the correlations scale with the inverse of the sample conductance,
whereas in the present work there is no small parameter that sets
the size of the long-range correlations, irrespective of sample
size or distance. In that sense, only the latter correlations are
truly long range.

This work was supported by the DFG, the NSF under Grant
No.\ DMR 0334499 and by the Packard foundation\,(PWB).
\bibliography{thesis,paperdef,paper,newpaper,book,refs}
\end{document}